\documentclass[12pt]{iopart}
\usepackage{graphicx}
\begin{document}

\title[Quantum Hall effect: the resistivity of 2D electron gas -
a thermodynamic approach]{Quantum Hall effect: the resistivity of
a two-dimensional electron gas - a thermodynamic approach}

\author{Maxim Cheremisin
\footnote[3]{To whom correspondence should be addressed
(maksim.vip1@pop.ioffe.rssi.ru)} }

\address{ A.F.Ioffe Physical-Technical Institute, St.Petersburg, Russia}

\begin{abstract}
Based on a thermodynamic approach, we have calculated the
resistivity of a 2D electron gas, assumed $dissipationless$ in a
strong quantum limit. Standard measurements, with extra current
leads, define the resistivity caused by a combination of Peltier
and Seebeck effects. The current causes heating(cooling) at the
first(second) sample contacts, due to the Peltier effect. The
contact temperatures are different. The measured voltage is equal
to the Peltier effect-induced thermoemf which is linear in
current. As a result, the resistivity is non-zero as $I\rightarrow
0$. The resistivity is a universal function of magnetic field and
temperature, expressed in fundamental units $h/e^{2}$. The
universal features of magnetotransport data observed in experiment
confirm our predictions.

\end{abstract}




\section{Introduction}
Understanding of electronic transport in solids subjected to high
magnetic fields was revised profoundly after the discovery of the
quantum Hall effect (Klitzing \etal 1980) for two-dimensional
electron gas( 2DEG ). This phenomenon manifests an extraordinary
transport behavior as the temperature approaches zero. The Hall
resistivity, $\rho _{xy}$, is quantized to $h/ie^{2}$ with $i$
being either an integer or rational fraction, while the
resistivity, $\rho _{xx}$, vanishes.

The main goal of this paper is to present a thermodynamic approach
to the IQHE problem. We address the problem of zero-temperature
transport, where all known scattering events are due to impurities
and long-range potential. Both cases were considered by Baskin
\etal (1978) and Fogler \etal (1997) classically, when 2DEG
conductivity was found to follow Drude formalism and vanished at a
magnetic field exceeding a certain value, dependent on the
dominant scattering. Moreover, the exact solution for 2DEG in a
strong magnetic field in the presence of impurities was found by
Baskin \etal (1978). For an impurity concentration per unit area,
$\mathcal{N}$, 2DEG conductivity vanishes when
$\mathcal{N}\mathbf{<}\Gamma =1/2\pi l_{B}^{2}$, where $\Gamma$ is
a zero-width Landau level(LL) density of states(DOS), and
$l_{B}=(\hbar c/eB)^{1/2}$ the magnetic length. Following the
above scenario, we further investigate the extreme quantum limit,
when 2DEG is considered $dissipationless$, and both the
conductivity and resistivity tensors are off-diagonal. The
thermodynamics of reversible processes are known to be valid
(Obraztsov 1964) in such a case. Our support of this idea is
confirmed by thermopower measurements data (Obloh \etal 1984),
found to agree with theoretical predictions for dissipationless
2DEG.

The crucial point of this paper is that an extraneous resistivity
can, nevertheless, arise from combined Peltier and Seebeck
thermoelectric effects (Kirby and Laubitz 1973, Cheremisin 2001).
The current leads to heating(cooling) at the first(second) sample
contacts, owing to the Peltier effect. The contact temperatures
are different, and the temperature gradient is linear in current.
The voltage drop is equal to the thermoemf induced by the Peltier
effect. As a result, the resistivity associated with the above
voltage is nonzero. We argue that this value may be identified as
the 2DEG resistivity proper. We compared our results with
experimental data. In fact, the possible undesirable influence of
thermoemf on QHE measurements was discussed among experimentalists
(Shahar, 2002).

\section{Analytical approach}
\begin{figure}
\begin{center}
\includegraphics[scale=0.6]{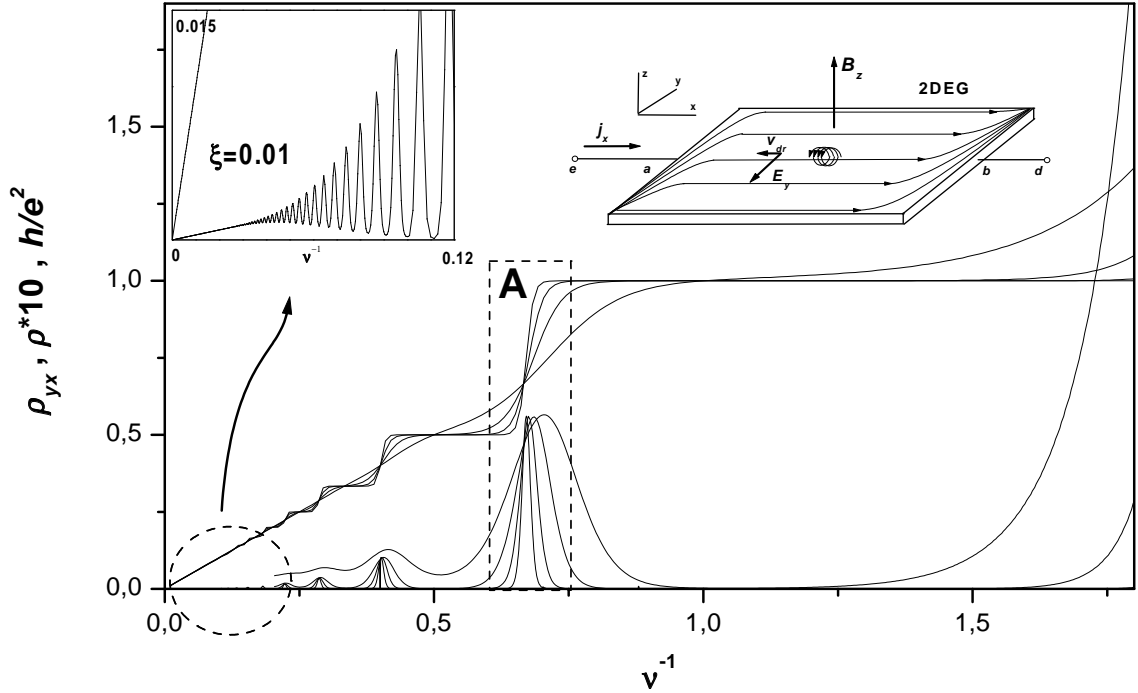}
\caption[]{\label{f.1}Magnetic field dependence of $\rho_{yx}$ and
$\rho $ (scaled by a factor of 10) given by (\ref{e.3}) for $\xi
=0.01,0.02,0.04,0.08$. The enlarged plot of data depicted in
section A is represented in figure \ref{f.2}. Insets: (left)
Low-field dependence of resistivities given by
(\ref{e.3},\ref{e.4});(right)QHE experimental setup}
\end{center}
\end{figure}

According to conventional thermodynamics, the chemical potential,
$\mu $, for system conductors+2DEG( see figure \ref{f.1}, inset)
is constant at equilibrium. Moreover, $\mu$ remains constant under
minor current induced fluctuations from the equilibrium. We
conclude that 2DEG is non-isolated. We argue that an external
reservoir of electrons, if it exists, could provide the pinning of
the 2DEG chemical potential. Baraff and Tsui(1980) and
Konstantinov \etal (1983) suggested that the remote ionized
donors(surface states in Si-MOS) may serve as such a reservoir. If
we suppose that $\mu$ is fixed, it can be shown that the 2DEG
density, $N$, changes with the magnetic field. Indeed, at
$T\rightarrow 0$, $N=i\Gamma $ electrons are required in order to
occupy $i$ levels. Hence, the Hall conductivity is given by
$\sigma _{yx}=Nec/B=ie^{2}/h.$ The reservoir must furnish the
opportunity for the variation of $N$ over sufficiently wide limits
in order to achieve observable $\sigma _{yx}$-plateaux.
Unfortunately, this scenario ceased (Klitzing \etal 1984), because
of insufficient strength of the reservoir in both models. In
contrast, the experiments (Nizhankovskii \etal 1986) support the
reservoir conception and point to an oscillatory dependence of
electron density on magnetic field. We claim that the 3D electrons
of the metal leads can play a role in such a reservoir. Indeed, in
a strong magnetic field, when only a few LLs are filled, the
absolute change in $N$ at a fixed $\mu $ is of the order of
density, $N_{0}=\frac{m\mu}{\pi\hbar ^{2}}$, of strongly
degenerated 2DEG at $B=0$. Hence, the number of 3D reservoir
electrons changes. The relative variation of the chemical
potential of 3D electrons is given by $\delta \mu /\mu
=\frac{2}{3}N_{0}lw/N_{3D}$, where $l,w$ are the length and width
of 2DEG, and $N_{3D}\approx 10^{23}$ is the total number of 3D
electrons. For a typical sample (1$\times $3 mm$^{2}$,
$N_{0}=10^{11}$cm$^{-2}$) we have $ \delta \mu /\mu \approx
10^{-15}$. Evidence shows that 2D chemical potential is fixed
under such conditions, providing the expected quantization
accuracy.

Let us consider a 2DEG in the x-y plane(figure \ref{f.1}, inset),
subjected to a magnetic field $B=B_{z}$. The 2DEG structure is
arbitrary, the electrons are supposed to occupy the first
size-quantization sub-band. For simplicity, we disregard spin
effects, the 2D electron energy spectrum being $\epsilon
_{n}=\hbar \omega _{c}(n+1/2)$, where $n=0,1..$ is the LL number,
and $\omega _{c}=eB/mc$ the cyclotron frequency. Then, we assume
no further LL broadening within the extreme quantum limit $\hbar
\omega_{c}>>kT$. A Hall-bar geometry sample is connected to the
current source by means of two identical leads. The contacts are
assumed to be ohmic. The voltage is measured between the open ends
("e" and "d"), kept at the temperature of the external thermal
reservoir. The sample is placed in a chamber with a mean
temperature $T_{0}$. Including the temperature gradient term, the
macroscopic current, $\mathbf{j}$, and the energy flux,
$\mathbf{q}$, densities are given by
\begin{equation}
\mathbf{j}=\mathbf{\sigma }^{\symbol{94}}(\mathbf{E-}\alpha \nabla
T),\quad \mathbf{q}=\left( \alpha T-\zeta /e\right)
\mathbf{j}-\mathbf{\kappa }^{\symbol{94}}\nabla T.
\label{e.1}
\end{equation}
Here, $\mathbf{E}=\nabla \zeta /e$ is the electric field, $\zeta
=\mu -e\varphi $ the electrochemical potential, and $\alpha $ the
thermopower. The conductivity, $\mathbf{\sigma }^{\symbol{94}}$,
and resistivity, $\mathbf{\rho }^{\symbol{94}}$, tensors are assumed
to be off-diagonal, therefore $\sigma _{yx}=1/\rho
_{yx}$. Then, $\mathbf{\kappa }^{\symbol{94}}=LT\mathbf{%
\sigma }^{\symbol{94}}$\ is the electron-related thermal
conductivity, and $ L=\frac{\pi ^{2}k^{2}}{3e^{2}}$ the Lorentz
number. Note that (\ref{e.1}) is valid for a confined-topology
sample, for which the diamagnetic surface currents (Obraztsov
1964) are accounted. Both Einstein and Onsager relationships are
satisfied. In fact, (\ref {e.1}) represents the current and energy
dissipationless fluxes caused by electron drift in crossed fields(
for 3D case, see Zyryanov 1964). Using (\ref{e.1}), we obtain
\begin{equation}
j=j_{x}=\sigma _{xy}E_{y},\quad j_{y}=\sigma _{yx}(E_{x}-\alpha \nabla
_{x}T)=0.
\label{e.2}
\end{equation}
The current represents the flux of electrons in crossed fields
with a drift velocity $v_{dr}=cE_{y}/B$, where $E_{y}$ is the Hall
electric field. The transverse electron flow caused by the
longitudinal electric field $E_{x}$ is compensated by that
associated with the downstream temperature gradient.

We can now determine the temperature gradient caused by the
Peltier effect. Recall that Peltier heat is generated by a current
crossing the contact between two different conductors. At that
contact (for example "a" in figure \ref{f.1}, inset), the
temperature, $T_{a}$, electrochemical potential $\zeta $, normal
components of the total current, $I$, and the total energy flux
are continuous. There exists a difference $ \Delta \alpha =\alpha
_{m}-\alpha $ between the thermopowers of the metal and 2DEG,
respectively. At $\Delta \alpha >0$, the charge crossing the
contact "a" gains an energy $e\Delta \alpha T_{a}$. Consequently,
$Q_{a}=I\Delta \alpha T_{a}$ is the amount of Peltier heat evolved
per unit time in the contact "a". For $\Delta \alpha >0$ and
current flow direction shown in figure \ref{f.1}, the contact "a"
is heated and the contact "b" is cooled. The contacts are at
different temperatures, and $\Delta T=T_{a}-T_{b}>0$. At small
currents, the temperature gradient is small and $T_{a,b}\approx
T_{0}$. In this case, $\alpha $ can be assumed to be constant,
and, hence, we disregard Thomson heating ($\sim IT\nabla \alpha $)
of 2DEG. As will be shown, the sample cooling caused by heat leak
via the leads and 3D substrate is negligible. Hence, we consider a
2DEG under adiabatic cooling conditions, when the amount of the
Peltier heat evolved at the contact "a" is equal to that absorbed
at the contact "b".

The energy flux is continuous at each contact, thus $\kappa
_{yx}\left. \nabla _{x}T\right| _{a,b}=-j\Delta \alpha T_{a,b}$.
Here, we take into account that the current is known to enter and
leave the sample at two diagonally opposite corners(figure
\ref{f.1}, inset). The temperature gradient is given by $\nabla
_{x}T=-j\Delta \alpha\rho _{yx}$, thus being linear in current.
The voltage drop, $U$, measured between the ends ''e'' and ''d''
is found to be $U=\int E_{x}dx=\int\alpha dT=\Delta \alpha
(T_{a}-T_{b})$. Ignoring conductor resistances, both the Hall and
Peltier-effect-induced resistivities yield
\begin{equation}
\rho = U/jl=s\rho _{yx},\quad \rho
_{yx}^{-1}=Nec/B=e^{2}/h\sum\limits_{n}f(\varepsilon _{n}),
\label{e.3}
\end{equation}
where $s=\frac{\alpha ^{2}}{L}$, $N=-\left( \frac{\partial \Omega
}{\partial \mu }\right) _{T}$ is the 2D density, $\Omega=-kT
\Gamma \sum\limits_{n}\ln \left[1+\exp \left(\frac{\mu
-\varepsilon _{n}}{kT}\right)\right]$ the thermodynamic potential,
and $f(\epsilon )$ the Fermi function. We take into account in
(\ref{e.3}) that for actual case of metal leads $\Delta \alpha
\simeq -\alpha $. Note that the dissipated power is positive,
since $\rho j^{2}>0$. (\ref{e.3}) can also then be applied to
four-probe Hall-bar measurements and to the 2DHG case.

\begin{figure}
\begin{center}
\includegraphics[scale=0.6]{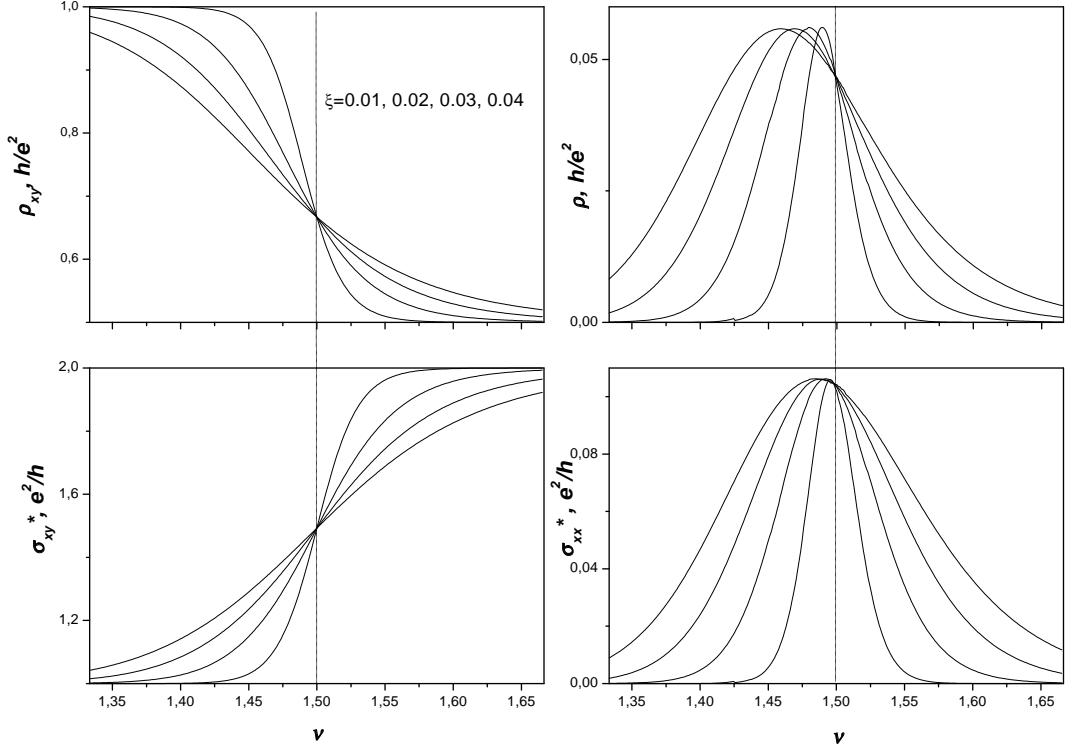}
\caption[]{\label{f.2}(top)The enlarged plot of $\rho_{yx},\rho $
data denoted by section A in figure \ref{f.1} and the related
conductivities(bottom) $\sigma _{yx}^{*} $, $\sigma _{xx}^{*}$
given by (\ref{e.5}) in the vicinity of half-filled LL $\nu
_{cr}=3/2$}
\end{center}
\end{figure}

Recall that in the strong quantum limit, 2D thermopower is a
universal quantity (Girvin and Jonson 1982), proportional to the
entropy per single electron: $\alpha =-{\frac{S}{eN}}$, where
$S=-\left( \frac{\partial \Omega }{\partial T}\right) _{\mu }$ is
the 2DEG entropy. It can be demonstrated that $S,N,\alpha$ are
universal functions of the reduced temperature $\xi =kT/\mu $ and
the magnetic field $\hbar \omega _{c}/\mu =\nu ^{-1}$, where $\nu
$ is the filling factor. Actually, $\nu $ corresponds to the
conventional filling factor, when spin effects are taken into
account. Using Lifshitz-Kosevich conventional formalism,
asymptotic formulae can easily be derived for $N,S$, and hence for
$\rho_{yx},\rho$, valid within low-temperature and magnetic field
limits $\nu ^{-1},\xi <1$:

\begin{equation}
\fl
N=N_{T}+\pi \xi N_{0}\sum\limits_{k=1}^{\infty
}\frac{(-1)^{k}\sin (2\pi k\nu )}{\sinh (r_{k})}, \quad
S=S_{0}-\pi ^{2}\xi kN_{0}\sum\limits_{k=1}^{\infty }(-1)^{k}\Phi
(r_{k})\cos (2\pi k\nu ), \label{e.4}
\end{equation}
where $ N_{T}=\frac{N_{0}}{2}\xi F_{0}(1/\xi)$ is the density,
$S_{0}= k\frac{N_{0}}{2}\frac{d}{d\xi }\left[\xi ^{2}F_{1}(1/\xi)
\right]$ the entropy at $B=0$. Then, $\Phi (z)=\frac{1-z\coth
(z)}{z\cdot \sinh (z)}$, $F_{n}(y)$ is the Fermi integral,
$r_{k}=2\pi ^{2}\xi\nu k$ the dimensionless parameter.

\begin{figure}
\begin{center}
\includegraphics[scale=0.6]{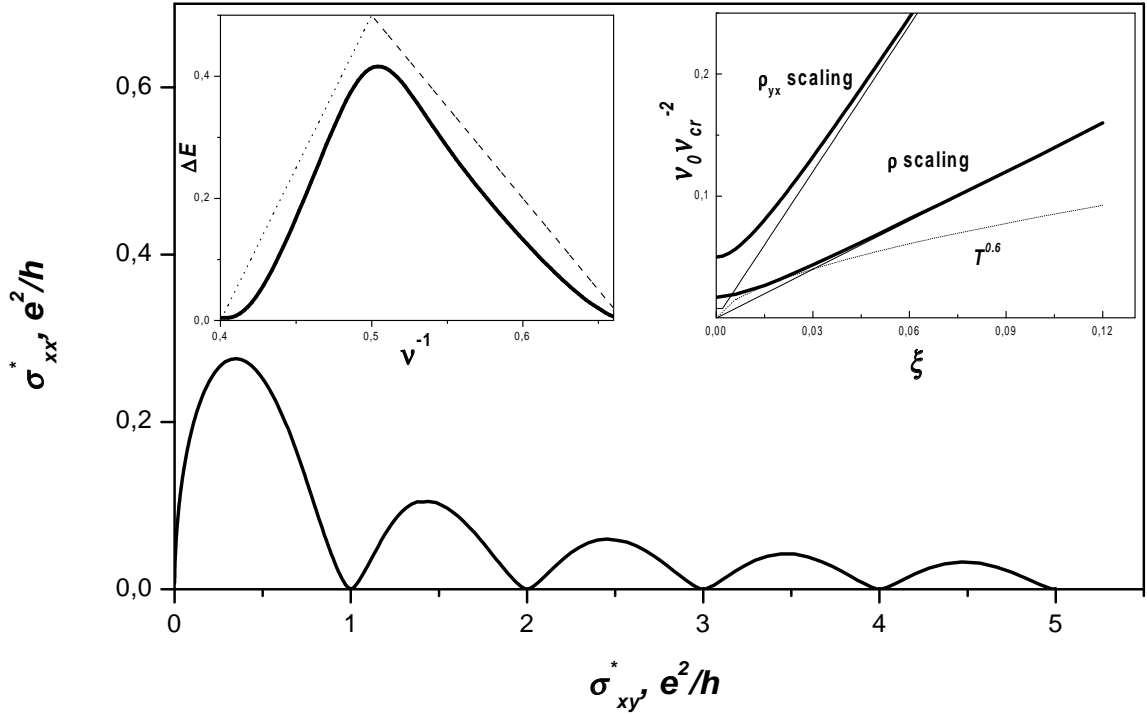}
\caption[]{\label{f.3} "Semicircle" plot: $\sigma _{xx}^{*}$ vs
$\sigma _{xy}^{*}$ at fixed temperature $\xi =0.01$. Insets:
(left) Activation energy of $\rho$-minimum in the vicinity $\nu=2$
at finite temperature $\xi=0.02$. Dashed(dotted)lines denote the
shift of n=1(2) LLs, with respect to Fermi energy; (right)
T-scaling of the logarithmic slope $\nu _{0}\nu _{cr}^{-2}$ for
both the Hall resistivity $\rho_{yx}$, and $\rho $ at $\nu
_{cr}=3/2$ and fixed LL broadening width $\gamma /\mu =0.04$. The
thin line corresponds to a zero LL width case. The dotted line
denotes power-law fit $T^{0.6}$.}
\end{center}
\end{figure}

According to the above reasoning, the resistivities given by
(\ref{e.3}) are universal functions(see figure \ref{f.1}) of
$\xi,\nu$ expressed in fundamental units $\frac{h}{e^2}$. Indeed,
$\alpha \sim k/e$, and, therefore $\rho \sim \rho _{yx}$. Within
classical magnetic field limits (figure \ref{f.1}, inset), we
obtain $\rho_{yx}=\frac{h}{\nu e^2}$. Then, in a strong magnetic
field when the chemical potential lies between the two LLs, $\rho
_{xy}$ is quantized to $\frac{h}{\nu e^2}$, while $\rho $ is
thermally activated with activation energy of the order of the
magnetic energy. Using (\ref{e.3}) we find out the activation
energy, $\Delta E=\frac{d\ln\rho}{d\xi^{-1}}$, for $\rho$ close to
filling factor $\nu=2$. The resonance-like curve( see figure
\ref{f.3},inset) is confined by two lines which define the shift
of the two neighboring LLs with respect to Fermi energy. At
$\nu=2$, the chemical potential lies in the middle of the two
proximate LLs, therefore the activation energy has a maximum
$\Delta E=0.415\sim 1/\nu=0.5$ close to an experimental (Wei \etal
1985) value of 0.42. With the help of (\ref{e.3}), at $\xi<<1$ for
integer fillings $\nu=1,2..$ the explicit formulae
$\rho=\frac{h}{e^2}\frac{3}{\pi^2
\nu^3}\left(2+\frac{1}{\nu\xi}\right)^2\exp\left(-\frac{1}{\nu\xi}\right)$
allows us to find out the above value of the activation energy.
The problem whether the conventional localization or present
mechanism dominate $\rho$-minima will be examined elsewhere.

\begin{table}
\caption{\label{tabl1}Critical resistivity within 2-1 and 1-0 QH
transition}

\begin{indented}
\lineup
\item[]
\begin{tabular}{@{}*{7}{l}}
\br \0\02D
gas&n($10^{10}cm^{-2}$)&$\mu$($10^{4}\frac{cm^2}{Vs}$)&$\nu_{cr},\rho^{cr}(2-1)$&$\nu_{cr},\rho^{cr}(1-0)$&paper\cr
\mr n-InP&33.7&\03.5&1.61/0.07&&Wei,1985 \cr
\chain&33&\03.5&1.53/0.19&&Wei,1988 \cr
\chain&20&\01.6&1.52/0.13&&Hwang,1993 \cr
\chain&\04&\09.4&\0\0\0\0\00.17&\0\0\0\0\0\0\0\0\00.88&Pan,1997
\cr \chain&\03&\03&&\0\0\0\00.56/1.18&Shahar,1998 \cr
n-GaAs&3.5-22.6&1.2-5.2&&0.55-0.67/0.77-1.32&Shahar,1995 \cr
\chain&22.7&\01.1&1.52/0.16&\0\0\0\00.62/1.18&Shahar,1997 \cr
\chain&20&80&1.60/0.11&\0\0\0\00.54/0.88&Wong,1997 \cr
p-SiC&34&\01.3&1.60/0.15&&Coleridge,1999 \cr
\chain&\08.7&&&\0\0\0\00.87/2.2&Hilke,1997 \cr \br
\end{tabular}
\end{indented}
\end{table}

At half-filled LLs ($\nu _{cr}=1/2,3/2...$) the $\alpha$, $\rho
_{yx}$ and $\rho $ approach the universal values $\alpha
_{cr}=-\frac{k}{e}\frac{\ln 2}{\nu _{cr}}$, $\rho
_{yx}^{cr}=\frac{h}{e^{2}}\nu _{cr}^{-1}$, and $\rho ^{cr}=\rho
_{yx}^{cr}\alpha _{cr}^{2}/L$, irrespective of temperature,
effective mass, etc. It can be demonstrated that, in the vicinity
of critical fillings, $\delta \nu =\nu -\nu _{cr}$, the Hall
resistivity can be represented as $\rho _{yx}=\rho
_{yx}^{cr}(1-\delta \nu /\nu _{0})$ for $\delta \nu /\nu _{0}<<1$.
Then, $\rho =\rho ^{cr}(1-3\delta \nu /\nu _{0})$, where $\nu
_{0}=4\nu _{cr}^{2}\xi $ is the logarithmic slope dependent on
sample and temperature. Note that $\nu _{0}\nu _{cr}^{-2}$ is
independent of the LL number. We argue that the above universal
values $\rho ^{cr}$, $\rho _{yx}^{cr}$ can be associated with the
"QH transition points" discussed in the press. To confirm this, in
figure \ref{f.2} the detailed dependencies $\rho (\nu ),\rho
_{yx}(\nu )$ in the vicinity of $\nu _{cr}=3/2$ are presented.
Evidence shows that this set of curves can be collapsed (Shahar
\etal 1997, Coleridge \etal 1999) into a single curve, since $\rho
,\rho _{yx}$ are universal functions of $\xi,\nu $. As an example,
at $\nu _{cr}=1.66$ (see Wei \etal 1985, spin is accounted) we
have $\rho^{cr}=0.034\frac{h}{e^2}$, consistent with an
experimental value of $0.07$. The experimental observations(see
Table\ref{tabl1}) demonstrate a certain range
$0.07-0.19\frac{h}{e^2}$ of $\rho^{cr}$-magnitude, which is higher
compared with that in theory. However, for 1-0 QH transition the
agreement between theory($\nu_{cr}=1/2,
\rho^{cr}=1.17\frac{h}{e^2}$) and experimental results is somewhat
better. We stress that "QH transitions" exhibit a certain
universality irrespective to sample, 2D density and $\sim
100$-fold change in scattering strength(see Table \ref{tabl1}),
hence strongly support the dissipation-free 2DEG scenario in
question.

It is worthwhile to mention that the existence of T-independent
"QH-transition points" is, in fact, the direct consequence of the
zero-width LL model in question. Indeed, for half-filled LL we
obtained the above results, since $\Omega =-kT\Gamma \ln 2$,
$S=k\Gamma \ln 2,$ then $N=\Gamma \nu_{cr}$. This apparent
discrepancy with respect to the second law of thermodynamics,
namely that $S\neq 0$ at $T=0$, will be discussed within the
realistic model of nonzero LLs width.

It is to be noted that the relationship between the resistivity
and conductivity of tensors components has not yet been
established experimentally. Within the QH plateaux, the Corbino
topology measurements (Dolgopolov \etal 1991) demonstrate that
$\sigma _{yx}=1/\rho _{yx}$. In practice, both the transverse and
the longitudinal conductivities are always derived from the
Hall-bar sample resistivity data. Assuming that both the $\rho
_{yx}$, $\rho $ components are those actually measured, we
calculate the related conductivities by means of transformation
$(\mathbf{\sigma }^{*})^{\symbol{94}}=(\mathbf{\rho
}^{\symbol{94}})^{-1}$ as follows:
\begin{equation}
\sigma _{yx}^{*}=\sigma _{yx}/(1+s^{2}),\quad \sigma_{xx}^{*}=s\sigma _{yx}^{*}.
\label{e.5}
\end{equation}
Figure \ref{f.2} represents $\sigma _{xx}^{*}{}(\nu ),\sigma
_{yx}^{*}(\nu )$ dependencies. These curves can be collapsed into
a single plot, as well as $\rho ,\rho _{yx}$(see above) in
consistent with experiments (Shahar \etal 1997, Coleridge \etal
1999). Using (\ref{e.5}), we can derive the relationship $(\sigma
_{xx}^{*})^{2}+(\sigma _{yx}^{*})^{2}=\sigma _{yx}^{*}\sigma
_{yx}$ between both the $\sigma _{xx}^{*}{},\sigma _{yx}^{*}$
components. This dependence(figure \ref{f.3}) is similar to that
known as the ''semicircle'' relation.

The known electron-phonon coupling is weak below $\sim $ 1K.
Hence, the heat transfer from 2DEG to the mixing chamber could
occur through the contacts and the leads connected to them.
However, the experiments (Mittal \etal 1994) at $B=0$ demonstrate
that the 2DEG alone is the dominant thermal resistance. The 2DEG
cooling is provided by electron thermal conductivity found to
follow Widemann-Franz law. Accordingly, our adiabatic cooling
approach is applicable. At the same time, remote black body
radiation heating is known to play an important role as
$T_{0}\rightarrow 0$. For example, the electron temperature can
differ (Mittal \etal 1994) from that of the mixing chamber
refrigerator(30mK) by as much as 100mK. The extraneous power
heating of electrons is caused by black body radiation( $\sim$pW )
from the walls of the vacuum can surrounding the refrigerator.
Hence, in ultra low-T experiments, 2DEG is overheated: the
electron temperature is higher than that of the bath, i.e.
$T>T_{0}$. Neglecting for a moment 2DEG overheating, we have
estimated an established temperature gradient. At $\nu _{cr}=5/2$,
we have $\alpha _{cr}=0.28k/e$, $\rho _{yx}^{cr}=2h/5e^{2}$. For a
$1\times 3$mm$^2$ sample and a current $I=2$nA, we obtain $\nabla
_{x}T\simeq \alpha _{cr}\rho _{yx}^{cr}I(Lw)^{-1}=16$mK/mm. At
liquid-helium temperatures, $\Delta T/T_{0}=0.01$, confirming our
approach as valid. However, at $\nu<<0.5$ our basic assumption of
a small temperature gradient is violated since $\alpha \rightarrow
\infty $. It is to be noted that at elevated currents, the local
temperature $T_{a}+\nabla _{x}T\,\cdot x$ is established in the
sample, dependent on the ratio $I/w$, and differing from $T_{0}$.
The thinner the sample and/or the higher the current, the greater
the difference from $T_{0}$. As a result, the $\rho$-peaks width
depends on the local temperature. This mechanism appears to play a
crucial role in current-polarity dependent QHE breakdown (Komiyama
\etal 1996).

We can now improve our model, assuming a LL broadening. DOS can be
represented as the sum of Gaussian peaks $ D(\varepsilon
)=\frac{\Gamma}{\gamma}
\sqrt{\frac{2}{\pi}}\sum\limits_{n}\exp \left[ -2\left( \frac{\varepsilon -\varepsilon _{n}%
}{\gamma }\right) ^{2}\right] $ with LL width, $\gamma $,
independent of the magnetic field. Obviously, our previous results
remain valid when $kT\gg $$\gamma $. In the opposite case of
low-temperature $kT\ll \gamma $, the LL broadening becomes
important and determines a $\rho_{yx}$-plateau width, $\rho $-peak
width and height. Indeed, at $kT\ll \gamma $, DOS can be
considered as constant in the vicinity of the n-th LL. A small
deviation of $\Delta =\mu -\varepsilon _{n}$, in the n-th LL, with
respect to chemical potential, results in a $\Gamma
\sqrt{\frac{2}{\pi}}\frac{\Delta}{\gamma }$ change in the number
of occupied states. For $\rho_{yx}$-scaling, the logarithmic slope
is $\nu _{0}=\sqrt{\frac{\pi}{2}}\frac{\nu_{cr}^2\gamma}{\mu}$,
i.e. it is proportional to LL width (see figure \ref{f.3}, inset).
Then, at $kT\ll \gamma $ the entropy yields $S_{cr}=\frac{ \pi
^{2}}{3}\Gamma \sqrt{\frac{2}{\pi }}\frac{k^{2}T}{\gamma }$, and
therefore the $\rho $-peaks height decreases at $T\rightarrow 0$.
In contrast, $\rho $-scaling data obeys the $\rho _{yx}$-scenario,
the logarithmic slope being $\nu_{0}/3$. Finally, at $T\rightarrow
0$ the height of the $\rho $ peaks decreases, but their width
remains finite.

We must emphasize that the effects related to nonzero LL can be
masked because of 2DEG overheating. Indeed, upon bath cooling, the
electron temperature would be constant, being controlled by remote
black body heat sources. In the case of strong heating, the
electron temperature is decoupled with respect to bath($T>T_{0}$),
and the condition $kT\ll \gamma $ cannot be achieved. As a result,
the logarithmic slope $\nu _{0}(T _{0})$ is at first linear, and
then saturates below a certain value of heat-dependent
temperature. Such a saturation behavior is similar to that
discussed above within the nonzero-width LL scenario. It should be
noted that, in both cases, one can attribute (see figure
\ref{f.3}, inset) the dependence $\nu _{0}(T _{0})$ to the power
law $T_{0}^{p}$, where $p<<1$. In actual fact, the question as to
whether the power law or linear scaling scenario is valid is still
the subject of heated debate(Huckestein 1995 , Sondhi \etal 1997).
Recent experiments (Shahar \etal 1998, Coleridge 1999, Balaban
\etal 1998) demonstrate that the linear dependence is preferable,
while power law application to restricted portions of data is
always possible. Neglecting LL broadening and 2DEG overheating,
for InGaAs/InP 2D electron gas( Shahar \etal 1998) at
$N_{0}=3\cdot 10^{10}$cm$^{-2}$ and $\nu _{cr}=0.58$(spin is
accounted), we obtain $4\nu _{cr}^{2}\xi /3=0.065\cdot T_{0}$(K)
for $\rho $-scaling. This result is consistent with the fit
$a\cdot T_{0}+b$, where $a=0.088$K$^{-1}$, $b=0.053$, as found by
Shahar \etal(1998). We attribute the non-zero logarithmic slope at
$T_{0}\rightarrow 0$ to an extraneous heating of 2DEG. Indeed,
$\rho(\nu)$-data exhibit (Shahar \etal 1998) a well-pronounced
"QH-transition", indicating the unimportance of LL broadening.

Finally, we address the question of 2DEG self-heating due to
finite energy dissipation $\rho j^{2}$. By equating the unit area power $\rho (I/w)^{2}$
dissipated in the sample (Koch \etal 1991) to that absorbed by
phonons (Chow \etal 1996) $P\sim T^{4}$, we can determine
the electron temperature which can be comparable with the bath temperature $%
T_{0} $ and $\gamma /k$. If $T>T_{0}$, we obtain the power-law fit
for a $\rho $-peak width, as $\Delta B\sim T^{p}\sim (I/w)^{p/2}$,
known for current and sample-width scaling. For $p=0.68$ (see Koch
\etal 1991), we have $\Delta B\sim w^{-0.34}$, consistent with
experimental finding $\Delta B\sim w^{-0.43}$. We argue that
simultaneous measurement (Balaban \etal 1998) of temperature and
frequency-dependent linear scaling $\Delta B\sim f,T $ points to
2DEG heating by incoming($\sim$ nW) ac power.

Up to now only the dc case has been discussed. However, our
approach can also be applied to the ac case. As shown by Kirby and
Laubitz(1973) and Cheremisin(2001), at $B=0$ the Peltier effect
related resistivity vanishes above a certain frequency, dependent
on thermal inertial processes. We recall that, in a quantizing
magnetic field, macroscopic current and heat fluxes are known to
depend on the magnetism of the conducting electrons(Obraztsov
1964). Actually, Eqs.(\ref{e.1}) describe the average current and
heat densities for a confined topology sample. We argue that the
diamagnetic currents could
define the dynamics of thermal processes in 2DEG. For $B=10$T , $%
m=0.068m_{0} $\ we have $\omega _{c}=2.8\cdot 10^{13}$Hz,
$l_{B}=70$A . For a sample size $l\sim 1$\ mm, the transit time of
an electron scattered at the sample boundary is $t\sim l(\omega
_{c}l_{B})^{-1}=5\cdot 10^{-9}c$. Accordingly, we can estimate the
critical frequency to be $f_{0}\sim 1/t=0.2$GHz. The smaller the
sample, the higher $f_{0}$. At the moment, however, a detailed
spectral analysis of $\rho $ is beyond the scope of the present
paper.

\section{Concluding remarks}
In conclusion, the magnetotransport measurements of 2D
electron(hole) gas, assumed to be dissipationless in the QH
regime, result in a resistivity associated with the Peltier and
Seebeck effects combined. This value is probably associated with
2D resistivity proper. The resistivity is a universal function of
magnetic field and temperature, expressed in units $h/e^{2}$. The
observed "QH transitions" demonstrate a certain universality
irrespective to sample, 2D density and $\sim 100$-fold change in
scattering strength, hence support our dissipationless scenario.
Our approach may be useful to eliminate the role of thermoelectric
effects within QHE regime.

\section*{Acknowledgments}
This work was supported by RFBR(grant 03-02-17588) and the
LSF(HPRI-CT-2001-00114, Weizmann Institute).

\section*{References}
\smallskip
\begin{harvard}
\item Balaban N Q, Meirav U and Bar-Joseph I 1998 {\it Phys.Rev.Lett.} {\bf 81} 4967
\item Baraff GA and Tsui D C 1981 {\it Phys.Rev.}B {\bf 24} 2274
\item Baskin E M, Magarill L I and Entin M V 1978 {\it Sov.Phys.JETP } {\bf 48} 365
\item Cheremisin M V 2001 {\it Sov.Phys.JETP} {\bf 92} 357
\item Chow E, Wei H P, Girvin S M, and Shayegan M  1996 {\it Phys.Rev.Lett.} {\bf 77} 1143
\item Coleridge P T, Zawadzki P 1999 {\it Sol. State Comm.} {\bf 112} 241
\item Coleridge P T 1999 {\it Phys. Rev.}B {\bf 60} 4493
\item Dolgopolov V T, Zhitenev N B and Shashkin A A 1991 {\it Europhys.Lett.} {\bf 14} 255
\item Fogler M M, Dobin A, Perel V I and Shklovskii B I 1997 {\it Phys. Rev.} B {\bf 56} 6823
\item Girvin S M and Jonson M 1982 {\it J.Phys.}C {\bf 15} L1147
\item Hilke M, Shahar D, Song S H, Tsui D C, Xie Y H and Don Monroe 1997 {\it Phys. Rev. B} {\bf 56} R15545
\item Hwang S W, Wei H P, Engel L W, Tsui D C, and Pruisken A M M 1993 {\it Phys.Rev.}B {\bf 48} 11416
\item Huckestein B 1995 {\it Rev. Mod. Phys.} {\bf 67} 357
\item Kirby C G M and Laubitz M J 1973 {\it Metrologia} {\bf 9} 103
\item Klitzing K von, Dorda G and Pepper M 1980 {\it Phys. Rev. Lett.}{\bf 45} 494
\item Klitzing K von and Ebert G 1983 {\it Physica(Utrecht)}B+C {\bf 117,118} 682
\item Koch S, Haug R J, Klitzing K von and Ploog K 1991 {\it Phys.Rev.Lett.} {\bf 67} 883
\item[] \dash In these experiments the total
current was kept constant(Koch S, private communication)
\item Komiyama S, Kawaguchi Y, Osada T and Shiraki Y 1996 {\it Phys.Rev.Lett.} {\bf 77} 558
\item Konstantinov O V, Merzin O A and Shik A Ya 1983 {\it Semoconductors } {\bf 17} 1073
\item Mittal A, Keller M W, Wheeler R G and Prober D E  1994 {\it Physica }B {\bf 194-196} 167
\item Nizhankovskii V I, Mokerov V G, Medvedev B K and Shaldin Yu.V 1986 {\it Sov.Phys.JETP} {\bf 63} 776
\item Obloh H and Klitzing K von and Ploog K 1984 {\it Surf.Sci.} {\bf 142} 236
\item Obraztsov Yu N 1964 {\it Sov. Phys. Solid State} {\bf 6} 331
\item Pan W, Shahar D, Tsui D C, Wei H P and Razeghi M 1997 {\it Phys. Rev.}B {\bf 55} 15431
\item Shahar D, Tsui D C, Shayegan M, Bhatt R N and Cunningham J E 1995 {\it Phys. Rev. Lett.} {\bf 74} 4511
\item Shahar D, Tsui D C, Shayegan M, Shimshoni E and Sondhi S L 1997 {\it Phys. Rev. Lett.} {\bf 79} 479
\item Shahar D, Hilke M, Li C C, Tsui D C, Sondhi S L and Razeghi M 1998 {\it Sol. State Comm. } {\bf 107} 19
\item Shahar D 2002, private communication
\item Sondhi S L, Girvin S M, Carni J P and Shahar D 1997 {\it Rev. Mod. Phys.} {\bf 69} 315
\item Wei H P, Chang A M, Tsui D C and Razeghi M 1985 {\it Phys. Rev.}B {\bf 32} 7016
\item Wei H P, Tsui D C, Paalanen M A and Pruisken A M M 1988 {\it Phys. Rev. Lett.} {\bf 61} 1294
\item Wei H P, Engel L W and Tsui D C 1994 {\it Phys. Rev.}B {\bf 50} 14609
\item Wong L W, Jiang H W and Schaff W J 1996 {\it Phys. Rev. B} {\bf 54} R17323
\item Zyryanov P S 1964 {\it Fizika Tverdogo Tela} {\bf 6} 3562

\end{harvard}
\smallskip

\end{document}